\documentclass[aps,twocolumn,groupedaddress,showpacs,nofootinbib]{revtex4-1}
\usepackage{graphicx}
\usepackage{amsmath}
\usepackage{amssymb}
\pdfoutput=1

\begin{document}
\bibliographystyle{prsty}
\title{Particle Production during Inflation in Light of PLANCK}

\author{Diana Battefeld}
\email{dbattefeld08@gmail.com}
\author{Thorsten Battefeld}
\email{tbattefe@gmail.com}
\author{Daniel Fiene}
\email{danielfiene@web.de}
\affiliation{Institute for Astrophysics, University of Goettingen, Friedrich Hund Platz 1, D-37077
Goettingen, Germany}
\date{\today}
\begin{abstract}
We consider trapped inflation in a higher dimensional field space: particle production at a dense distribution of extra species points leads to a terminal velocity at which inflation can be driven in steep potentials. We compute an additional, nearly scale invariant contribution to the power-spectrum, caused by back-scattering of the continuously produced particles. Since this contribution has a blue tilt, it has to be sub-dominant, leading to an upper bound on the coupling constant between the inflatons and the extra species particles. The remaining allowed parameter space is narrow.
\end{abstract}
\maketitle

\section{Introduction}
The observation of the cosmic microwave background radiation by the PLANCK satellite is consistent with the predictions of simple, small-field slow-roll models: perturbations are adiabatic, Gaussian, with a red spectral index and gravitational waves\footnote{The BICEP2 experiment \cite{Ade:2014xna} claims a detection of $r\sim \mathcal{O}(0.1)$, which would rule out small field models if confirmed independently.} are not observed \cite{Ade:2013zuv,Ade:2013uln,Ade:2013ydc}.

On the other hand, inflationary models in string theory are commonly of the multi-field type, with models on landscapes  (higher dimensional moduli spaces \cite{Bousso:2000xa,Susskind:2003kw,Douglas:2006es}) appearing to be common. Extracting generic predictions of inflation on such landscapes has received increased attention, as evident in the ongoing investigation of random landscapes \cite{Aazami:2005jf,Frazer:2011tg,Frazer:2011br,McAllister:2012am,Dias:2012nf,Battefeld:2012qx,Pedro:2013nda,Battefeld:2013xwa,Marsh:2013qca} the effect of decaying fields during inflation (staggered inflation) \cite{  Battefeld:2008py,Battefeld:2008ur,Battefeld:2008qg,Battefeld:2010rf,Battefeld:2010vr}, cascade inflation \cite{Ashoorioon:2006wc,Ashoorioon:2008qr} or multi-field open inflation \cite{Sugimura:2011tk,Battefeld:2013xka,Sugimura:2013cra} among others multi-field models \cite{Dimopoulos:2005ac,Easther:2005zr,Becker:2005sg,Majumdar:2003kd,Tye:2008ef,Ashoorioon:2009wa,Ashoorioon:2009sr,Ashoorioon:2011ki,Ashoorioon:2011aa}. It should be noted that the presence of eternal inflation, requiring the choice of a measure as well as anthropic reasoning (see e.g.~\cite{Olum:2012bn,Schiffrin:2012zf} for a recent analysis of the measure problem and \cite{Freivogel:2011eg} for a review of proposed measures), can hamper solid predictions. While PLANCK is consistent with single field models, it does not preclude, but merely constrains most multi-field effects, see e.g.~\cite{Elliston:2013afa} for a summary. For instance,  the absence of primordial non-Gaussianities \cite{Ade:2013ydc} curtails the curvature of the end-of-inflation hypersurface \cite{Bernardeau:2002jf,Lyth:2005qk,Salem:2005nd,Alabidi:2006wa,Bernardeau:2007xi,Sasaki:2008uc,Naruko:2008sq,Battefeld:2013xwa,Elliston:2013afa}, modulated reheating \cite{Dvali:2003em,Zaldarriaga:2003my,Kofman:2003nx,Vernizzi:2003vs,Bernardeau:2004zz,Ichikawa:2008ne,Battefeld:2007st,Byrnes:2008zz,Suyama:2007bg}, curvatons \cite{Enqvist:2001zp,Lyth:2001nq, Moroi:2001ct,Enqvist:2005pg,Linde:2005yw,Malik:2006pm,Sasaki:2006kq}, modulated trapping \cite{Battefeld:2011yj,Langlois:2009jp} and particle production during inflation \cite{Barnaby:2010ke,Barnaby:2010sq,Battefeld:2011yj,  Barnaby:2010vf,Barnaby:2011pe}, among other effects, see \cite{Byrnes:2010em,Suyama:2010uj} for reviews.
 Furthermore, the absence of gravitational waves \cite{Ade:2013zuv} puts pressure on large-field models ($r\sim 0.1$ due to the Lyth bound \cite{Lyth:1996im}), whether they are of the single- or multi-field type.

In this article we consider trapped inflation in higher dimensional field spaces \cite{Battefeld:2010sw}, which is a class of multi-field models operating on moduli spaces containing a dense distribution of extra-species points (ESPs). 

ESPs are locations at which additional degrees of freedom become light and can be produced kinematically; once produced, they need to be incorporated into the low energy effective field theory. ESPs are generic in string theory and often associated with additional symmetries, see \cite{Seiberg:1994rs,Seiberg:1994aj,Witten:1995im,Intriligator:1995au,Strominger:1995cz,Witten:1995ex,Katz:1996ht,Bershadsky:1996nh,Witten:1995gx} for some examples (see also \cite{Bagger:1997dv,Watson:2004aq} for a discussion of the string Higgs effect). Applications include moduli trapping \cite{Kofman:2004yc,Watson:2004aq,Patil:2004zp,Patil:2005fi,Battefeld:2005av,Battefeld:2005wv,Cremonini:2006sx,Greene:2007sa},
   effects on inflation \cite{Chung:1999ve,Elgaroy:2003hp,Romano:2008rr}, trapped inflation \cite{Kofman:2004yc,Green:2009ds,Silverstein:2008sg,Battefeld:2010sw} (see also \cite{Bueno Sanchez:2006eq,Bueno Sanchez:2006ah,Brax:2009hd,Matsuda:2010sm,Brax:2011si,Lee:2011fj,Moroi:2013tea}),
  modulated trapping \cite{Langlois:2009jp,Battefeld:2011yj,D'Amico:2013iaa}, and preheating
   \cite{Kofman:1997yn,Battefeld:2006cn,Battefeld:2012wa}
    (see \cite{Bassett:2005xm} for a review and extensive references on preheating)\footnote{See \cite{Allahverdi:2011aj} for particle production at a point of enhanced gauge symmetry after inflation within the MSSM \cite{Allahverdi:2006iq,Allahverdi:2006cx,Allahverdi:2006we}.} . ESP-distributions can be dense, as in trapped inflation \cite{Kofman:2004yc,Green:2009ds,Silverstein:2008sg,Battefeld:2010sw}\footnote{Monodromy inflation \cite{Silverstein:2008sg} is a realization of trapped inflation \cite{Kofman:2004yc,Green:2009ds}.}, or sparse, as in \cite{Battefeld:2011yj}, depending on the moduli space under consideration. In this paper we model the extra species particles phenomenologically by including additional scalar fields that are coupled quadratically to the inflatons, as in \cite{Kofman:2004yc} (see \cite{Barnaby:2010vf,Barnaby:2011qe,Barnaby:2012tk} for the incorporation of gauge fields).

Trapped inflation is based on the inclusion of backreaction of the extra species particles onto inflatons \cite{Kofman:2004yc,Green:2009ds,Silverstein:2008sg,Battefeld:2010sw}: once particles are produced, an attractive force towards the ESP results, affecting inflationary dynamics \cite{Kofman:2004yc}. If the dimensionality of field space is large, $D\gg 1$, and ESPs are dense, a terminal velocity\footnote{This terminal velocity at weak coupling should not be confused with the speed limit at strong coupling \cite{Silverstein:2003hf} leading to DBI-inflation \cite{Alishahiha:2004eh}. } $v_{t}\sim gx^2$ results \cite{Battefeld:2010sw}. $v_{\mathrm{t}}$ becomes independent of the potential in the large $D$ limit. As long as the potential is steep enough, the trajectory is traversed at this constant speed; thus, functional fine tuning, i.e., the $\eta$ problem, is relaxed. Preheating was discussed in \cite{Battefeld:2012wa}, entailing qualitatively new resonance effects: efficient preheating is likely for dense distributions, $x\lesssim  0.001$, in contrast to preheating with a single ESP at the VEV of the inflatons \cite{Battefeld:2008bu,Battefeld:2008rd,Battefeld:2009xw,Braden:2010wd}.

Intrinsic perturbations (i.e., solutions to the homogeneous equation for perturbations) are expected to have a red spectral index and a suppressed amplitude similar to perturbation in single field trapped inflation \cite{Green:2009ds}. We leave the computation of this spectrum to a future publication in lieu of focussing on additional curvature fluctuations sourced by backscattering of the produced particles of the inflaton condensate. This effect was computed for distinguishable ESP encounters in a series of papers by Barnaby et al.~\cite{Barnaby:2009mc,Barnaby:2009dd,Barnaby:2010ke,Barnaby:2010sq,Barnaby:2010vf,Barnaby:2011pe,Barnaby:2011qe,Barnaby:2012tk,Barnaby:2012xt}: in the power-spectrum, a bump (with a suppressed ringing pattern) results at a wave-number corresponding to the Hubble radius at the time of the encounter. In our setup, several ESPs are encountered at any given time, leading to a superposition of bumps. In Sec.~\ref{sec:powerback}, we compute analytically a nearly scale-invariant  contribution to the power-spectrum, $P_{\mathrm{bs}}$,  resulting from this superposition. We find a small blue tilt that is ruled out by PLANCK (a red index was found at the $5\sigma$ level). Consequently, the amplitude of $P_{\mathrm{bs}}$ needs to be sub-dominant, leading to an upper bound on the coupling constant, $g\lesssim \mathcal{O}(0.01)$. 

We discuss the implication for trapped inflation in Sec.~\ref{sec:discussion}, concluding that only a narrow region of parameter space remains viable. We comment on non-Gaussianities in Sec.~\ref{sec:nonG}, which receive an additional contribution from the same effect. We expect a similar constraint on $g$ from bounds on the bi-spectrum, but leave a computation to future work. 

The detailed outline of this paper is as follows: we start with a brief recap of trapped inflation in Sec.~\ref{sec:recap}, providing a heuristic argument motivating the presence of a terminal velocity. The conditions to drive trapped inflation at $v_{\mathrm{t}}$ are summarized in Sec.~\ref{sec:conditions}. Readers familiar with \cite{Battefeld:2010sw} may skip these sections. 
 After commenting briefly on the intrinsic power-spectrum in Sec.~\ref{sec:intrinsic}, we derive the additional contribution from backscattering in Sec.~\ref{sec:powerback}, followed by a discussion of the allowed values of $g$ and $x$ in Sec.~\ref{sec:discussion}. We comment on the effectiveness of preheating and non-Gaussianities thereafter, before concluding in Sec.~\ref{sec:conslusion}.
  
Throughout this article we set the reduced Planck mass to unity,
\begin{eqnarray}
M_P^2= \frac{1}{8\pi G}\equiv 1 \,.
\end{eqnarray}

\section{Brief Recap: Trapped Inflation in a Higher Dimensional Field Space \label{sec:recap}}
In \cite{Battefeld:2010sw} the mechanism of trapped inflation \cite{Kofman:2004yc,Green:2009ds,Silverstein:2008sg} has been generalized to higher dimensional field spaces and preheating has been discussed in \cite{Battefeld:2012wa}. We refer the interested reader to \cite{Battefeld:2010sw} for a thorough derivation of the results stated in this section, but would like to provide a heuristic argument to aid the reader's intuition.

\subsection{A Heuristic Argument \label{sec:heuristic}}
Extra species points (ESPs) are locations in field space at which additional degrees of freedom become light, so that they can be produced kinematically.
 Here, we wish to consider a higher dimensional field space with potential $V(\varphi_i)$, $i=1\dots D$, $D\gg 1$ and a dense, homogeneous and isotropic \footnote{The assumption of homogeneity and isotropy is made to simplify computations, and not derived from any underlying principle.} ESP distribution, characterised by an average inter-ESP distance $x$. We model the extra degree of freedom at a single ESP at $\vec{\varphi}^{\mbox{\tiny ESP}}$ by an additional light scalar field coupled to the inflaton via the interaction Lagrangian
\begin{eqnarray}
\mathcal{L}_{\mathrm{int}}=-\frac{g^2}{2}\chi^2\sum_{i=1}^{D}(\varphi_i-\varphi_i^{\mbox{\tiny ESP}})^2\,. \label{Lint}
\end{eqnarray}
For simplicity, we assume canonical kinetic terms and an identical coupling $g$ between the inflatons and the extra fields. If the bare mass $m_\chi$ is small and \cite{Kofman:1997yn,Barnaby:2009mc}
\begin{eqnarray}
g>\frac{H^2}{v} \,, \label{lowerboundg}
\end{eqnarray}
with $v\equiv |\dot{\vec{\varphi}}|$ and $H=\dot{a}/a$ the Hubble parameter,
particle production during an ESP encounter can be computed analytically, identical to particle production during preheating \cite{Kofman:1997yn}. The produced particle density depends sensitively on the distance of closest approach, i.e.~the impact parameter 
\begin{eqnarray}
\mu=\mbox{min}(|\vec{\varphi}(t)-\vec{\varphi}_{\mbox{\tiny ESP}}|)\,,
\end{eqnarray}
so that  
\begin{eqnarray}
\rho_\chi\propto e^{-\frac{g}{v}\mu^2}\,.\label{rhochi} \label{suppression}
\end{eqnarray} 
Evidently, if $\mu>\sqrt{v/g}$, particle production is suppressed and this particular ESP can be neglected. However, if the trajectory comes closer, a fraction of the inflatons' kinetic energy is transferred to $\chi$-particles. As the trajectory moves away from the ESP, these particles become heavy and, via the coupling in (\ref{Lint}), lead to an attractive force towards the ESP that diminishes over time in an expanding universe, $\rho_\chi\propto a^{-3}$. Such a single encounter can slow down, bend and, for strong coupling, temporarily trap the trajectory \cite{Kofman:2004yc,Battefeld:2011yj}. While interesting effects similar to the curvaton scenario can result for such a single encounter \cite{Battefeld:2011yj}, we wish to consider the other extreme of a dense ESP distribution, so that individual encounters become indistinguishable from each other.

Due to the exponential suppression in (\ref{suppression}), only ESPs in a cylinder around the trajectory with radius of order $\sqrt{v/g}$ need to be considered. If $\sqrt{v/g}\lesssim x$, ESPs are encountered individually as in \cite{Battefeld:2011yj}, if at all. On the other hand, if $\sqrt{v/g}>x$, many ESPs are close enough to the trajectory at any given time for particle production to take place. Since we distribute ESPs homogeneously, no particular direction is singled out. Hence, if the trajectory in field space is sufficiently straight in the absence of ESPs, the resulting backreaction of the produced $\chi$ particles onto the inflatons is opposite to the velocity $\dot{\vec{\varphi}}$.

\begin{figure}[tb]
\includegraphics[scale=0.45,angle=0]{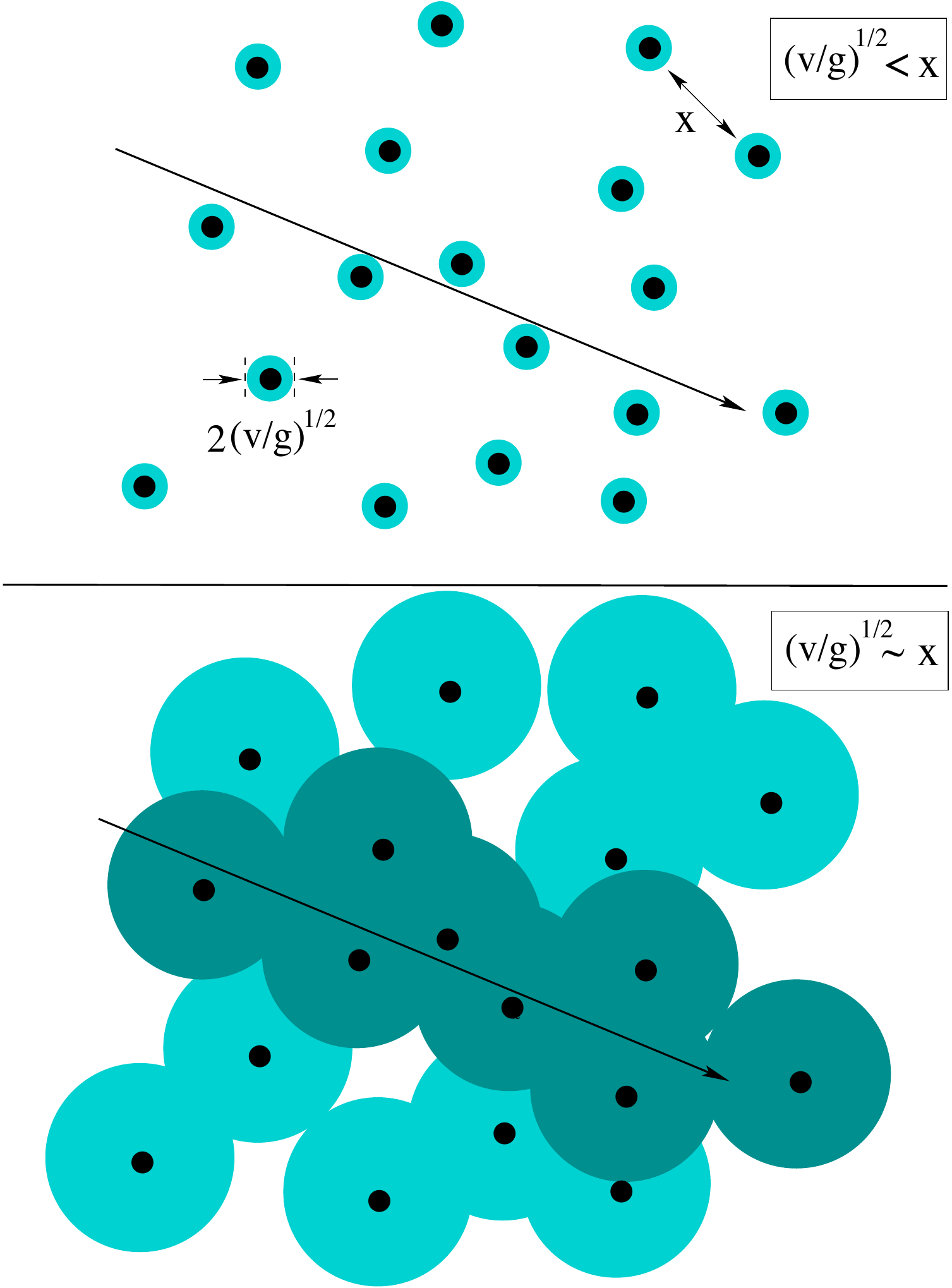}
   \caption{Schematic: ESPs (black dots) with average inter-ESP distance $x$ distributed on a $D=2$ dimensional field space; the arrow denotes an approximately straight trajectory given by $V(\vec{\phi})$. The blue circles have radius $\sqrt{v/g}$; only those ESP that have an imact parameter of $\mu\sim \sqrt{v/g}$ (dark blue) can affect inflationary dynamics, due to the exponential suppression in (\ref{rhochi}). Top: the speed in field space $v$ is small, so that $\mu > \sqrt{v/g}$ for almost all ESPs. Except for rare encounters with single ESPs, no particle production takes place. Bottom: as $v$ increases, $\sqrt{v/g}\sim x$, and many ESPs become within reach of the trajectory. Back-reaction provides a friction force,  leading to a movement at the terminal velocity $v_{\mathrm{t}}$ in (\ref{approximatevt}). 
    \label{pic:1}}
\end{figure}

Let's consider inflatons placed somewhere high up in a steep potential $V$: initially at rest, the fields accelerate due to the gradient. As long as $v<gx^2$, hardly any ESPs are in reach and the inflatons keep accelerating. Once $v\sim gx^2$ is reached, particles at many ESPs are produced, resulting in an opposing force to the speed, and thus the gradient of the potential if the trajectory is sufficiently straight. This force becomes exceedingly strong as $v$ increases, subsequently reducing $v$. On the other hand, if $v$ dips below $gx^2$, the speed picks up again due to the gradient of the potential. Thus, after initial oscillations, balancing these two forces leads to a movement at a terminal velocity of order $gx^2$, see Fig.~\ref{pic:1} for an illustration.

This heuristic argument can be made precise, see \cite{Battefeld:2010sw}, leading to the terminal velocity 
\begin{eqnarray}
v_{\mathrm{t}}&=&gx^2\Delta\,,\\
\Delta&\equiv& \left(\frac{(2\pi)^3 3H}{g^5x^4}\frac{\partial V}{\partial \phi}\right)^{2/(D+4)}\label{defDelta}
\end{eqnarray}
in the large $D$ limit (this result ignores factors of order $1/D$), where we defined $\phi$ as the field along the trajectory. Since $\Delta \rightarrow 1$ for large $D$, we can approximate 
\begin{eqnarray}
v_{\mathrm{t}}\approx gx^2 \label{approximatevt}
\end{eqnarray} 
in the large $D$ limit, which is independent of the potential (assuming that the slope of the potential is steep enough to drive $v$ towards $v_{\mathrm{t}}$). 

\subsection{Conditions for Inflation at the Terminal Velocity \label{sec:conditions}}

In order for (\ref{approximatevt}) to be self-consistent and inflation to be driven at the terminal velocity, several conditions need to be satisfied (see \cite{Battefeld:2010sw} for more details):
\begin{enumerate}
\item The potential needs to be steep enough to ignore Hubble friction, yielding the condition $v_{\mathrm{t}}\ll v_{\mathrm{SR}}\approx |\partial V/ \partial \phi|/(3H)$. If $v_{\mathrm{t}} > v_{\mathrm{SR}}$, the slow roll regime is entered before the terminal velocity is reached.
\item To avoid prolonged oscillations around $v_{\mathrm{t}}$, the time for backreaction to act needs to be smaller than the characteristic time needed to change the speed due to the potential. This entails the lower bound $D\gg 2\ln(|\partial V/\partial \phi|^2(2\pi)^3/(gx)^6)$.
\item The trajectory should not be strongly curved over a few Hubble times, so that backreaction is approximately anti-parallel to the velocity and the gradient, yielding the upper bound $v_{\mathrm{t}}\lesssim 10^{-4}$.
\item The potential energy needs to dominate over the kinetic energy which in turn needs to be bigger than the energy in produced particles, to guarantee that inflation takes place.
\item At least $N=60$ e-folds of inflation are needed to solve the standard problems of the big bang. 
\end{enumerate}

\section{Contributions to the Power-spectrum \label{sec:intrinsic}}
The study of perturbations in trapped inflation is subtle, since backreaction of $\chi$ particles onto the background is included at the level of the equations of motion, not the Lagrangian. In the one dimensional case, an  analysis of perturbations in $\phi$ in the uniform curvature gauge for a deSitter background is given in \cite{Green:2009ds}, which can be generalized to our case. This computation takes place in two steps: first, the solution of the homogeneous equation leads to the intrinsic power-spectrum. In v1 of the this article on the arxiv, a discussion of the intrinsic power-spectrum was included, following the standard approach to cosmological perturbation theory as used in \cite{Battefeld:2010sw}. This computation, as well as the section pertaining to perturbations in \cite {Battefeld:2010sw}, is not applicable, since it is based on the standard Lagrangian for cosmological perturbations; instead, the methods employed in \cite{Green:2009ds} should be used. We plan to correct this computation in a forthcoming separate publication and focus on a second contribution to the power-spectrum induced by the peculiar solution to the inhomogeneous equation, i.e.~ by the inclusion of a source term due to backscattering: perturbations in the produced particle density (responsible for the speed limit and thus crucial for the background solution) back-scatter off the inflaton condensate. For a single ESP encounter, this back-scattering effect leads to a bump in the power-spectrum, accompanied by a (suppressed) ringing pattern  \cite{Barnaby:2009mc,Barnaby:2009dd}. A dense superposition of these bumps should lead to a nearly scale invariant contribution to the power-spectrum, that can reach the COBE limit if $g$ is sufficiently large, as mentioned in \cite{Battefeld:2010sw}. We expect the this second contribution to dominate over the intrinsic power-spectrum without tuning, in line with the results of \cite{Green:2009ds}.

The computation of this contribution and its scalar spectral index is the aim of this paper. We will show that this contribution $P_{\mathrm{bs}}$ carries an observationally ruled out blue spectral index so that it must be sub-dominant, leading to an upper bound on $g$. Thus, the intrinsic power-spectrum has to saturate the COBE bound while satisfying observational constraints on the scalar spectral index and the tensor to scalar ratio  or an additional mechanism has to be invoked to provide the power-spectrum. 

In the former case, fields move at the terminal velocity while scales in the observational window leave the horizon; a regular slow roll regime is entered thereafter. This setup can be valid for complicated potentials as long as $V$ is steep enough and the conditions in Sec.~\ref{sec:conditions} are satisfied. If the spectral index or amplitude fail to meet observational requirements one may use a curvaton \cite{Enqvist:2001zp,Lyth:2001nq,Moroi:2001ct} or modulated reheating \cite{Dvali:2003em,Zaldarriaga:2003my,Kofman:2003nx,Vernizzi:2003vs},  to provide the power-spectrum. Such an addition would render the model rather complicated and one may argue to discard it based on Ockhams razor.

If trapped inflation is not operational during the last sixty e-folds of inflation, it may be present and useful at earlier times: if fields start out high up in a steep potential on some complicated (random) landscape in string theory inflation is rare in the absence of a speed limit; if it occurs, it usually takes place near a saddle point \cite{Aazami:2005jf,Frazer:2011tg,Frazer:2011br,McAllister:2012am,Battefeld:2012qx,Marsh:2013qca}. If a speed limit is present, inflation is already taking place when the saddle is encountered, avoiding the over-shoot problem and addressing the initial value problem of inflationary cosmology (why is $\dot{\phi}^2 /2 \ll V$?). In this setup, the initial inflationary phase at the terminal velocity takes place before scales relevant for observations left the horizon and has therefore no observable impact. 


\section{The Power-spectrum from Back-scattering \label{sec:powerback}}

 \begin{figure}[tb]
\includegraphics[scale=0.4,angle=0]{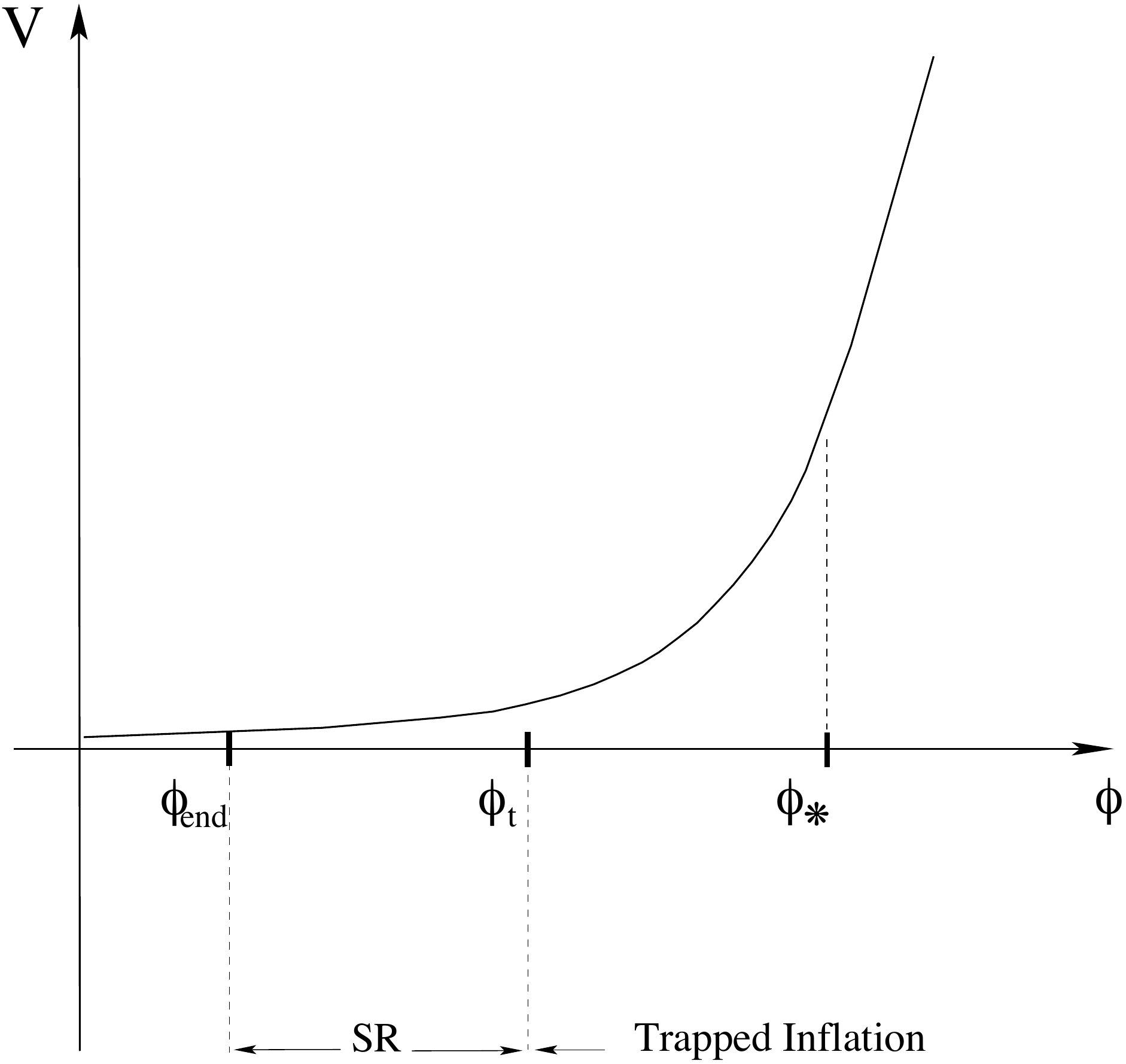}
   \caption{An inflationary potential ($s>2$), denoting the region of slow roll dynamics for $\phi_{\mathrm{end}}<\phi<\phi_{\mathrm{t}}$ and trapped inflation at the terminal velocity $v_{\mathrm{t}}$ in (\ref{approximatevt}) for $\phi>\phi_{\mathrm{t}}$, with the transitional field value $\phi_{\mathrm{t}}$ in (\ref{defphit}). \label{pic:2}}
\end{figure}
In \cite{Barnaby:2009mc,Barnaby:2009dd} the effect of back-scattering of a single ESP onto fluctuations of a single inflaton field were computed, yielding a bump-like contribution to the power-spectrum that can be approximated by
\begin{eqnarray}
P_{\mbox{\tiny ESP}}\approx A\left(\frac{\pi e}{3}\right)^{3/2}\left(\frac{k}{k_{\mbox{\tiny ESP}}}\right)^3\exp\left(-\frac{\pi}{2}\frac{k^2}{k_{\mbox{\tiny ESP}}^2}\right)\,,
\end{eqnarray}
which ignores a ringing pattern in the large $k$ tail (we are not interested in these oscillations, since they will be averaged out via the superposition of bumps). The amplitude was computed in \cite{Barnaby:2009dd} by comparison to lattice field simulations to
\begin{eqnarray}
A\approx  10^{-6}g^{15/4}\,, \label{AESP}
\end{eqnarray}
 which was tested for a quadratic potential and  $g^2=1,0.1,0.01$; it was found to be a good approximation up to factors of order unity \cite{Barnaby:2009dd}. We shall use this estimate to put constraints on $g$ subsequently, keeping this theoretical uncertainty in mind. The location of the peak is set by the Hubble scale at the ESP-encounter, 
\begin{eqnarray}
k_{\mbox{\tiny ESP}}=cH_{\mbox{\tiny ESP}} e^{N{\mbox{\tiny ESP}}}
\end{eqnarray}
where 
\begin{eqnarray}
N_{\mbox{\tiny ESP}}\equiv  \int_{\phi_*}^{\phi_{\mbox{\tiny ESP}}}\frac{H(\phi)}{\dot{\phi}}\, \mbox{d}\phi
\end{eqnarray}
is the number of e-folds between $\phi_*$ at $N=60$ e-folds before the end of inflation and the ESP encounter at $\phi_{\mbox{\tiny ESP}}$, and the proportionality constant is given by $c=\sqrt{gv}/H_{\mbox{\tiny ESP}}$ \cite{Barnaby:2009mc,Barnaby:2009dd} (note that our notation differs from the one in \cite{Barnaby:2009mc,Barnaby:2009dd}).

For us, there are two important differences compared to \cite{Barnaby:2009mc,Barnaby:2009dd}: firstly, while $v$ is given by the slow roll speed in \cite{Barnaby:2009mc,Barnaby:2009dd}, it is the terminal velocity $v=v_{\mathrm{t}}=gx^2$ for us. As a consequence, we get $c=gx/H$ so that
\begin{eqnarray}
k_{\mbox{\tiny ESP}}=gx e^{N_{\mbox{\tiny ESP}}}\,.\label{kESP}
\end{eqnarray}
with 
\begin{eqnarray}
N_{\mbox{\tiny ESP}}&\approx &-\frac{1}{v_{\mathrm{t}}}\int_{\phi_*}^{\phi_{\mbox{\tiny ESP}}}H(\phi)\,\mbox{d}\phi\\
&\approx& N(1-y^{1+s/2})\,,\label{NESPofy}
\end{eqnarray}
where we defined
\begin{eqnarray}
y\equiv \frac{\phi_{\mbox{\tiny ESP}}}{\phi_*}
\end{eqnarray}
and used the monomial potential
\begin{eqnarray}
V=\sum_{i=1}^D\frac{\lambda}{s}\varphi_i^s=\frac{\lambda}{s}\phi^s\,,\label{potential}
\end{eqnarray}
with $s\geq 2$ in the last step as an explicit example. We only consider ESP encounters in the trapped inflation regime, i.e.~$\phi_{ESP}>\phi_{\mathrm{t}}$; the latter results by equating the slow roll speed 
\begin{eqnarray}
v_{\mathrm{SR}}=\sqrt{\frac{\lambda s}{3}}\phi^{s/2-1}
\end{eqnarray}
with the terminal velocity $v_{\mathrm{t}}$, so that the transitional field value becomes
\begin{eqnarray}
\phi_{\mathrm{t}}\equiv \left(v_{\mathrm{t}}\sqrt{\frac{3}{\lambda s}}\right)^{2/(s-2)}\,,\label{defphit}
\end{eqnarray}
for $s>2$ (for $s=2$ the slow roll speed is constant, so that either slow roll or movement at at the terminal velocity is operational during the inflationary regime).
Thus, the slow roll regime is operational for
\begin{eqnarray}
\phi_{\mathrm{end}}\lesssim \phi \lesssim\phi_{t}\,,
\end{eqnarray}
with $\phi_{\mathrm{end}}=s/\sqrt{2}$ at which the potential slow roll parameter equals one, while inflation is driven at the terminal velocity for $\phi>\phi_{t}$, see Fig.~\ref{pic:2} for an illustration. The transition between the two regimes is smooth\footnote{When $v_{\mathrm{SR}}$ dips below $v_{t}$, particle productions ceases, and the effect of the already produced particles red-shifts away since $\rho_\chi\propto a^{-3}$. Hence, the velocity slowly decreases from $v_{\mathrm{t}}$ tracking $v_{\mathrm{SR}}$.} so that no sharp features in observables are expected. The total number of e-folds 
\begin{eqnarray}
N&=&\int_{\phi_{\mathrm{t}}}^{\phi_{\mathrm{end}}} \frac{H}{\dot{\phi}}\,\mbox{d}\phi+\int_{\phi_*}^{\phi_{t}} \frac{H}{v_{\mathrm{t}}}\,\mbox{d}\phi \label{fullN}\\
&\equiv& N_{\mathrm{SR}}+N_{\mathrm{t}}
\end{eqnarray}
can be computed to
\begin{eqnarray}
N_{\mathrm{SR}}&\approx& \frac{1}{2s}\left(\phi_{\mathrm{t}}^2-\phi_{\mathrm{end}}^2\right)\,,\label{NSR}\\ 
N_{t}&\approx& \sqrt{\frac{\lambda}{3s}}\frac{1}{v_{\mathrm{t}}}\frac{2}{s+2}\left(\phi_*^{1+s/2}-\phi_{\mathrm{t}}^{1+s/2}\right)\,,
\end{eqnarray}
for $s>2$, where we used the slow roll approximation for $N_{\mathrm{SR}}$ and $v_{\mathrm{t}}=\mbox{const}$ as well as $3H^2\simeq V$ for $N_{\mathrm{t}}$.  Since 
\begin{eqnarray}
\lambda(\phi_*)=3 sH_*^2\frac{1}{\phi_*^s}\,, \label{deflambda}
\end{eqnarray}
and $\phi_{\mathrm{t}}=\phi_{\mathrm{t}}(\lambda(\phi_*))$ via (\ref{defphit}), it is usually not possible to solve $N(\phi_*)=60$ for $\phi_*$ analytically, but once the inflationary energy scale $H_*$ is specified, it may always be done numerically.

\begin{figure}[tb]
\includegraphics[scale=0.7,angle=0]{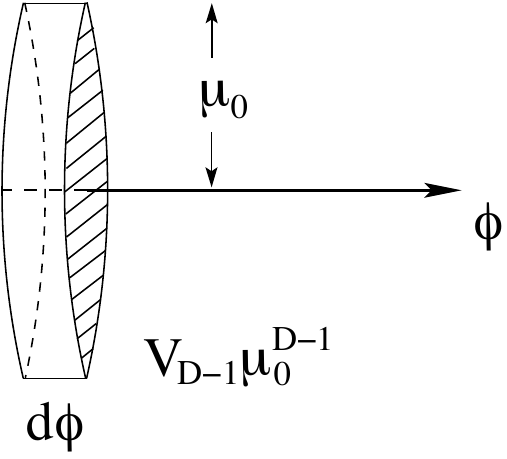}
   \caption{Schematic showing an infinitesimal volume element around the trajectory.  Only ESPs up to a distance of $\mu_0$ in (\ref{defmu0}) affect inflationary dynamics via backreaction and potentially curvature perturbations via back-scattering (see Fig.~\ref{pic:1}). \label{pic:3}}
\end{figure}

A second difference compared to \cite{Barnaby:2009mc,Barnaby:2009dd} is that ESPs are not encountered head on, but distributed around the trajectory and passed in rapid succession. As a result, particle production at a single ESP is suppressed by an impact parameter $\mu$, i.e.~, the distance of closest approach to the ESP, leading to $n_\chi\propto \exp({-g\mu^2/v})$, which carries over to the energy density in (\ref{rhochi}) and all associated effects. These  consequences were computed in \cite{Battefeld:2010sw}, where it was shown that particle production, and thus backreaction, is dominated by ESPs with impact parameters close to 
\begin{eqnarray}
\mu_0=x\sqrt{\frac{D}{2\pi}}\,; \label{defmu0}
\end{eqnarray} 
particle production of ESPs further away is exponentially suppressed, while ESPs closer to the trajectory are too few to affect the dynamics strongly if $D$ is large; the latter is a direct consequence of the fact that most of the volume, and thus ESPs, in a hypersphere is close to its surface in the large D limit. Thus, the corresponding particle density $n_\chi$ carries a suppression factor of $\exp{(-D/2)}$, which in turn carries directly over to $P_{\mbox{\tiny ESP}}$; we may thus write the cumulative contribution to the power-spectrum from back-scattering\footnote{For trapped inflation in one dimension, the corresponding spectrum from back-scattering was computed in \cite{Green:2009ds} by a different approximation scheme.} as
\begin{eqnarray}
P_{\mathrm{bs}}=\int_{\mbox{\tiny traj.}}\mbox{d}n_{\mbox{\tiny ESP}}P_{\mbox{\tiny ESP}}\,,
\end{eqnarray}
with 
\begin{eqnarray}
\nonumber P_{\mbox{\tiny ESP}}\approx A\left(\frac{\pi e}{3}\right)^{3/2}\left(\frac{k}{k_{\mbox{\tiny ESP}}}\right)^3\exp\left(-\frac{\pi}{2}\frac{k^2}{k_{\mbox{\tiny ESP}}^2}-\frac{D}{2}\right)\,,\\
\end{eqnarray}
and we defined the effective ESP line-density
\begin{eqnarray}
\mbox{d}n_{\mbox{\tiny ESP}}\equiv V_{D-1}\frac{\mu_0^{D-1}}{x^D}\mbox{d}\phi\,,
\end{eqnarray}
where $V_{D-1}$ is the volume of a $D-1$ dimensional unit-sphere. See Fig.~\ref{pic:3} for a schematic of the infinitesimal volume element to be integrated along the trajectory.

Replacing $\phi$ by $y$ as the integration variable and defining the dimensionless function
\begin{eqnarray}
\nonumber &&F(k)\equiv \int_0^{\infty}\left(\frac{k}{k_{\mbox{\tiny ESP}}(y)}\right)^3\exp\left(-\frac{\pi}{2}\left(\frac{k}{k_{\mbox{\tiny ESP}}(y)}\right)^2\right)\mbox{d}y\\
&&
\end{eqnarray}
we get
\begin{eqnarray}
P_{\mathrm{bs}}\approx \left(\frac{\pi e}{3}\right)^{3/2}A_{\mbox{\tiny ESP}}\frac{\phi_*}{x}F(k)\,,
\end{eqnarray}
were we used (\ref{defmu0}) so that
\begin{eqnarray}
V_{D-1}\frac{\mu_0^{D-1}}{x^D}\approx \frac{e^{D/2}}{x}\,,\label{VD-1}
\end{eqnarray}
and extracted the leading order result in a $1/D$ expansion. To be concrete, we used $V_{D}=\pi^{D/2}/\Gamma(1+D/2)$ and the large argument limit of the $\Gamma$-function.  Note that the suppression factor of $e^{-D/2}$ of a single ESP encounter with impact parameter $\mu_0$ is compensated by the large number of traversed ESPs.

Replacing $y$ by $k_{\mbox{\tiny ESP}}$ via (\ref{kESP}) and (\ref{NESPofy}),
and defining
\begin{eqnarray}
z&\equiv& \ln\left(\frac{k}{H_*}\right)\,,\\
\kappa&\equiv& \frac{k_{\mbox{\tiny ESP}}}{H_*}\,,
\end{eqnarray}
we can write the integral explicitly as
\begin{eqnarray}
\nonumber F(z)&=&\int_0^{gxe^N/H_*}\frac{2}{s+2}\left(1-\frac{1}{N}\ln\left(\frac{\kappa H_*}{gx}\right)\right)^{-s/(s+2)}\\
 &&\frac{1}{N\kappa}\left(\frac{e^z}{\kappa}\right)^{3}\exp\left(-\frac{\pi}{2}\left(\frac{e^z}{\kappa}\right)^2\right)\,\mbox{d}\kappa\\
 &\equiv&\int_0^{gxe^N/H_*} f(\kappa,z) \,\mbox{d}\kappa\,.
\end{eqnarray}
To compute the scalar spectral index of the additional contribution to the power-spectrum,
\begin{eqnarray}
n_s^{\mathrm{bs}}-1\equiv  \frac{\mbox{d}\ln P_{\mathrm{bs}}}{\mbox{d} \ln k}\bigg|_*=\frac{1}{F} \frac{\mbox{d} F}{\mbox{d}z}\bigg|_{z=0}\,,
\end{eqnarray}
we need $F$ and its derivative at $z=0$. 

Let us focus on $F(0)$ first: the integrand peaks at
\begin{eqnarray}
\bar{\kappa}(z)\equiv \frac{\sqrt{\pi}}{2}e^z\,,
\end{eqnarray}
so that we may approximate the logarithm in $f(\kappa,z)$ by its value at $\bar{\kappa}$. The remaining integral can be approximated by
\begin{eqnarray}
F(0)\approx \frac{1}{\pi N}\frac{\sqrt{2}}{s+2}\left(1-\frac{1}{N}\ln\left(\frac{\sqrt{\pi}H_*}{2gx}\right)\right)^{-s/(s+2)}\!\!\!\!\!\! .
\end{eqnarray}
Similarly, we can compute $\mbox{d}F/\mbox{d}z|_{z=0}$ by first approximating the logarithm in $f(\kappa,z)$ by its value at $\bar{\kappa}(z)$, computing the partial derivative of the entire integrand with respect to $z$, approximating the remaining elementary integrals and finally setting $z=0$, yielding after some algebra
\begin{eqnarray}
\nonumber \frac{\mbox{d} F}{\mbox{d}z}\bigg|_{z=0}\!\!\!&\approx &\frac{\sqrt{2}s}{\pi N^2(s+2)^2}\left(1-\frac{1}{N}\ln\left(\frac{\sqrt{\pi} H_*}{2gx}\right)\right)^{-2\frac{s+1}{s+2}}.\\
&&
\end{eqnarray}
We checked the analytic approximations numerically, and we found them to be accurate at the percent level. With these approximations, the scalar spectral index simplifies to
\begin{eqnarray}
n_s^{\mathrm{bs}}-1\approx \frac{s}{s+2}\frac{1}{N-\ln\left(\frac{\sqrt{\pi} H_*}{2gx}\right)}\,.\label{nbsfinal}
\end{eqnarray}
Evidently, this additional contribution to the power-spectrum has a blue spectral index $n_s^{\mathrm{bs}}-1\sim 10^{-2}$ for all $s$ (the logarithm gives a factor of order one for the allowed values of $g$ in (\ref{allowedintervallg})), which is ruled out by the PLANCK satellite at the $5\sigma$-level. Hence
\begin{eqnarray}
\nonumber P_{\mathrm{bs}}\!&\approx&\! \frac{A}{ N}\left(\frac{ e}{3}\right)^{\frac{3}{2}}\frac{\phi_*}{\sqrt{v_{\mathrm{t}}}} \frac{\sqrt{2\pi g}}{s+2}
\left(1-\frac{1}{N}\ln\left(\frac{H_*\sqrt{\pi}}{2\sqrt{v_{\mathrm{t}} g}}\right)\right)^{-\frac{s}{s+2}}\!\!\!\!\!\!\!\!\!\!\!,\\
&&\label{Pbsfinal}
\end{eqnarray}
where we used $x=\sqrt{v_{\mathrm{t}}/g}$, 
can't provide the dominant contribution to the power-spectrum, but has to satisfy
\begin{eqnarray}
P_{\mathrm{bs}}\ll P_{\zeta}\,.
\end{eqnarray}

\subsection{Discussion: Consequences of the Allowed Values of $g$ and $x$ \label{sec:discussion}}

Demanding $P_{\mathrm{bs}} \lesssim 0.1 P_{\zeta}$, yields the upper bound  $g \lesssim g_{\mathrm{max}}$ by using $P_{\mathrm{bs}}$ from (\ref{Pbsfinal}), the approximate amplitude $A$ from (\ref{AESP}) and $\phi_*$ from solving $N(\phi_*)=60$. This bound depends on the inflationary energy scale $H_*$ and the terminal velocity $v_{\mathrm{t}}$. If the intrinsic power-spectrum is computed, these parameters can be expressed in terms of observables such as the scalar spectral index and the tensor to scalar ration. 

Fortunately, for reasonable\footnote{Taking e.g.~$H_*\sim 10^{-5}$ assumes an inflationary energy scale in line with slow roll models where $(H_*^2/(2\pi\dot{\phi}))^2=P_{\zeta}\sim 2 \times 10^{-9}$ is set by the Cobe normalization (using $v_{\mathrm{t}}\sim 10^{-6}$ as in the estimate that follow). This slow roll result {\emph {does not}} apply for trapped inflation at the terminal velocity, which requires the computation of the intrinsic power-spectrum as in \cite{Green:2009ds} (not provided in this paper). If $H_*$ were considerably below $10^{-5}$, the lower bound on $g$ in (\ref{allowedintervallg}) would be relaxed, but the upper bound would hardly be affected due to the logarithmic dependence.} values of $H_*$ the logarithm in (\ref{Pbsfinal}) can be ignored, so that
\begin{eqnarray}
g_{\mathrm{max}}&\approx& \left(10^5 \frac{P_\zeta \sqrt{v_{\mathrm{t}}}}{\phi_*}\left(\frac{3}{e}\right)^{\frac{3}{2}}\frac{s+2}{\sqrt{2\pi}}\right)^{4/17}
\label{gmax}\,.
\end{eqnarray}
As $v_{\mathrm{t}}$ has to be below the slow roll speed to be relevant, $\phi_*$ needs to be below the value it would have if ESPs were absent. Thus, taking $s=3$ and $\phi_*= 19$ as a concrete example, we get
\begin{eqnarray}
g\lesssim g_{\mathrm{max}} \approx 0.022\, v_{\mathrm{t}}^{2/17}\,. 
\end{eqnarray}
Taking $v_{\mathrm{t}}\sim 10^{-6}$ or smaller so that $v_{\mathrm{t}}$ is below $v_{\mathrm{SR}}$ yields $g \lesssim 0.044 \sim \mathcal{O}(10^{-2})$ as an upper bound. Along with the lower bound on $g$ in (\ref{lowerboundg}), $g_{\mathrm{min}}\approx H_*^2/v_{\mathrm t} \sim 10^{-4}$ if $v_{\mathrm t}\sim 10^{-6}$ and $H_*\sim 10^{-5}$ during inflation at the terminal velocity,
 we arrive at the allowed interval
\begin{eqnarray}
\mathcal{O}(10^{-4}) \lesssim  g \lesssim  \mathcal{O}(10^{-2})\,.\label{allowedintervallg}
\end{eqnarray}
The upper bound on $g$ is comparable to the one originating from a single bump in the power-spectrum in \cite{Barnaby:2009dd}, and somewhat lower than the estimate in \cite{Battefeld:2010sw}. Whereas individual bumps may be used to match outliers/oscillating features in the power-spectrum, and thus improve a fit, a seizable contribution of $P_{\mathrm{bs}}$ is not desirable due to its blue tilt.

The interval for $g$ is not particularly wide, but allowed values do not appear to be overly fine tuned either and cover the same range commonly considered for studies of preheating after inflation \cite{Dolgov:1989us,Traschen:1990sw,Kofman:1997yn}. In fact, the fields at ESPs in the vicinity of the final resting place can act as preheat matter fields, as investigated in \cite{Battefeld:2012wa}: there, it was found that preheating is qualitatively different than in models that have an ESP at the VEV of the inflatons; while de-phasing of inflatons tends to suppress parametric resonance \cite{Battefeld:2008bu,Battefeld:2008rd,Battefeld:2009xw,Braden:2010wd} (see \cite{Bassett:1997gb,Bassett:1998yd,Battefeld:2009xw}  for the case of two inflations, that still permits resonances), two new effects leading to efficient preheating were found: particle production during the first in-fall can already comprise a seizable energy transfer if the trapped inflation regime lasts until preheating commences, but it is never complete. Subsequent broad resonance is generically suppressed due to de-phasing of the fields, but if an ESP happens to lie at a well defined distance from the VEV, a prolonged narrow resonance regime can complete preheating. In \cite{Battefeld:2012wa} it was concluded that both effects are important/likely if the average inter ESP distance is of order $x\sim 10^{-3}$ or smaller ($g=0.5 \times 10^{-4}$ was used). 
Since $x=\sqrt{v_{\mathrm{t}}/g}$ the allowed interval of $g$ corresponds to 
\begin{eqnarray}
\mathcal{O}(0.01)\lesssim x \lesssim \mathcal{O}(0.1)\,. \label{intervallx}
\end{eqnarray}
Such values of the average inter ESP separation appear reasonable in light of the known examples of moduli-spaces in string theory. 
Since $x\sim 10^{-3}$ lies outside of this interval, it is unlikely for an ESP to be at the right position for a prolonged narrow resonance regime. Furthermore, particle production during the first in-fall does not take place if the slow roll regime is operational towards the end of inflation. The latter is the case for monomial potentials if $\phi_{\mathrm t}$ in (\ref{defphit}) is bigger than $\phi_{\mathrm{end}}$. However, if the inflationary model is a large-field one, one may/should expect $x$ to differ by a factor of order one over the course of inflation, which might boost preheating; if preheating is absent, fields either decay via tachyonic instabilities \cite{Felder:2000hj,Dufaux:2006ee,Battefeld:2009xw} or the standard theory of reheating.

To summarize, while a small value of $P_\zeta$ still requires fine tuning of the inflationary energy scale\footnote{A computation of the intrinsic power-spectrum (not presented here) is needed to check if the amplitude of $\mathcal{P}_{\mathrm{int}}$ can be large enough to saturate the COBE bound; if this is the case, one can relate $P_\zeta$ to $H_*$ and tune $H_*$ accordingly. },  the functional fine tuning of the potential needed for large-field models in the absence of additional symmetries is absent for inflation at the terminal velocity, since inflation is due to a small $v_{\mathrm t}$, which in turn can be achieved by reasonable values of the coupling constant $g$ and inter ESP-distance $x$; the main requirement is a steep enough potential to drive $\dot{\phi}$ towards the terminal velocity, i.e.~,$v_{\mathrm{SR}}>v_{\mathrm{t}}$ (see Sec.~\ref{sec:conditions} for additional, mild conditions). In this regime, we showed that the additional contribution to the power-spectrum from backscattering is observationally ruled out by PLANCK, since it caries a blue spectral index. As a consequence, only the narrow interval in (\ref{allowedintervallg}) remains viable for trapped inflation at the terminal velocity, severely reducing the motivation to consider trapped inflation in the first place.

\subsubsection{Non-Gaussianities \label{sec:nonG}}
Let us comment briefly on non-Gaussianities:  conclusions of \cite{ Battefeld:2006sz,Battefeld:2009ym,Elliston:2011dr,Elliston:2011et,Elliston:2012wm},
 we do not expect large intrinsic non-Gaussianities to be generated during inflation even if the potential is more structured than in the simple monomial cases considered here, since the trajectory has to be reasonable straight (see point 3 in Sec.~\ref{sec:conditions}). However, another source of non-Gaussianities is present: as back-scattering leads to an additional contribution to the power-spectrum, it also acts as a source for higher order correlation functions, such as the bi-spectrum. In \cite{Barnaby:2010ke}, the shape-function due to a distinguishable ESP encounter was computed (see also \cite{Barnaby:2010sq}), entailing a localised feature as well as oscillatory components. Such a highly structured bi-spectrum (two integrals can't be expressed in closed form) is difficult to compare with observations: non-linearity parameters $f_{NL}$ for smooth shapes (local, equilateral, orthogonal, etc.) are essentially blind to localized and/or oscillatory features, and even modal expansions \cite{Fergusson:2009nv,Meerburg:2010ks,Ade:2013ydc} reach their limit of applicability fast for such oscillating signals, as discussed in \cite{Battefeld:2011ut}. However, in our case many copies of the shape function in \cite{Barnaby:2010ke} are superimposed, similar to the individual bumps in the power-spectrum. Hence, a smooth, nearly scale invariant bi-spectrum results, which should be much simpler to constrain. It would be interesting to perform this computation and compare the resulting upper bound on $g$ with the one stemming from the power-spectrum (we expect them to be comparable). We leave this interesting project to a future study.

\section{Conclusion \label{sec:conslusion}}

The presence of extra species points (ESPs) on moduli spaces in string theory is a generic phenomenon. If fields evolve on such a higher dimensional landscape ($D\gg 1$) with a dense distribution of ESPs, a speed limit at the terminal velocity $v_{\mathrm{t}}~\sim gx^2$ is present ($g$ is the coupling constant to the extra species particles and $x$ the average inter-ESP distance), potentially leading to trapped inflation.

We investigated the feasibility of this inflationary scenario in light of PLANCK's observation of a red-spectral index, $n_s=0.9603\pm 0.0073$. In addition to the intrinsic curvature perturbation which has a suppressed amplitude and is expected to carry a red spectral index (not computed here), another contribution to the power-spectrum is present in trapped inflation: the produced extra species particles backscatter off the inflaton condensate, leading to additional curvature fluctuations. A single ESP encounter leads to a bump in the power-spectrum at a wave-number set by the Hubble radius at the time of the event, as computed by Barnaby et.al.. Since the terminal velocity is due to the superposition of many ESP encounters at any given time, the resulting superposition of bumps leads to an additional, nearly scale invariant contribution to the power-spectrum, $P_{\mathrm{bs}}$. We computed analytically the amplitude (\ref{Pbsfinal}) and the scalar spectral index (\ref{nbsfinal}) of $P_{\mathrm{bs}}$. The index is always blue and therefore observationally ruled out by PLANCK. As a consequence, $P_{\mathrm{bs}}$ needs to be sub-dominant, which  leads to an upper bound on the coupling constant, $\mathcal{O}(10^{-4}) \lesssim g \lesssim \mathcal{O}(10^{-2})$ for $v_\mathrm{t}\sim 10^{-6}$ (the lower bound is required by the employed method to compute particle production at ESPs; it may be relaxed if the inflationary energy scale turns out to be considerably lower in trapped inflation compared to slow roll models, while still consistent with the COBE normalization.); correspondingly, the average inter-ESP distance needs to satisfy $\mathcal{O}(0.01) \lesssim x \lesssim \mathcal{O}(0.1)$. Values for $g$ and $x$ in these narrow intervals provide the remaining parameter space for trapped inflation at a terminal velocity. 

Given these narrow ranges, the appeal of trapped inflation is reduced as fine-tuning re-emerges to suppress the contribution to the power-spectrum from backscattering. Further, it is not clear if observational constraints can be met in this narrow range: the intrinsic power-spectrum needs to be computed to check whether or not the amplitude and spectral index of the power-spectrum as well as the tensor to scalar ratio are consistent with observations\footnote{The results in v1 of this paper on the arxiv pertaining to the intrinsic power-spectrum were faulty, due to the unjustified use of a standard Lagrangian for cosmological perturbations. An improved version of this computation is in progress by two of the authors (D.B. and T.B.) of this article.}.  Further, the lower bound on $x$ indicates that preheating at ESPs after inflation is most likely ineffective; hence, inflatons decay either via tachyonic instabilities or the standard theory of reheating. 
An additional contribution to non-Gaussianities from backscattering is present, but given that $P_{\mathrm{bs}}$ needs to be sub-dominant, we expect these non-Gaussianities to be below observational bounds.

\begin{acknowledgments}
We would like to thank L.~Lorentz for initial collaboration and N.~Barnaby for discussions. T.~B. and D.~B. would like to thank the APC (Paris) for hospitality and D.~Langlois for support.
\end{acknowledgments}

\end{document}